\documentclass[%
 aip,
 amsmath,amssymb,
 reprint,%
]{revtex4-1}

\usepackage{graphicx}% Include figure files
\usepackage{dcolumn}% Align table columns on decimal point
\usepackage{bm}% bold math
\usepackage{tablefootnote}
\usepackage[utf8]{inputenc}
\usepackage[T1]{fontenc}
\usepackage{mathptmx}
\usepackage{etoolbox}
\usepackage{verbatim}

\makeatletter
\def\@email#1#2{%
 \endgroup
 \patchcmd{\titleblock@produce}
  {\frontmatter@RRAPformat}
  {\frontmatter@RRAPformat{\produce@RRAP{*#1\href{mailto:#2}{#2}}}\frontmatter@RRAPformat}
  {}{}
}%
\makeatother
\begin{document}

\preprint{AIP/123-QED}

\title[An interpretable molecular descriptor for machine
learning predictions in atmospheric science]{An interpretable molecular descriptor for machine
learning predictions in atmospheric science\\}
% Force line breaks with \\
\author{L. Lind}
 \affiliation{Department of Applied Physics, Aalto University, FI-00076 Aalto, Finland.}%Lines break automatically or can be forced with \\
\author{H. Sandstr\"om}%
   \affiliation{Physics Department, TUM School of Natural Sciences, Technical University of Munich, 85748 Garching b. München, Germany}
       \affiliation{Department of Applied Physics, Aalto University,  FI-00076 Aalto, Finland.}
   \affiliation{Atomistic Modeling Center, Munich Data Science Institute, Technical University of Munich, 85748 Garching b. München, Germany}%Lines break 

\author{P. Rinke}%
  \affiliation{Physics Department, TUM School of Natural Sciences, Technical University of Munich, 85748 Garching b. München, Germany}
      \affiliation{Department of Applied Physics, Aalto University, FI-00076 Aalto, Finland.}

      \email{hilda.sandstroem@tum.de}
   
   \affiliation{Atomistic Modeling Center, Munich Data Science Institute, Technical University of Munich, 85748 Garching b. München, Germany}%Lines break 
   \affiliation{Munich Center for Machine learning, 80333 Munich, Germany}%Lines break 

%\affiliation{
%s%\\This line break forced with \textbackslash\textbackslash
%}%

\date{\today}% It is always \today, today,

\begin{abstract}
The study of aerosol formation and chemistry using machine learning is limited by the lack of molecular descriptors suited to atmospheric compounds. Interpretable models are particularly affected because they often rely on dictionary-based descriptors tied to specific molecular substructures, which currently fail to capture the full range of organic atmospheric compounds, including large, highly oxidized molecules common in the atmosphere. We introduce ATMOMACCS, an interpretable descriptor combining the 166 binary keys of the MACCS fingerprint with motifs inspired by the SIMPOL method for estimating saturation vapor pressures. We show that ATMOMACCS based models improve predictions of saturation vapor pressures (7-8\% error reduction), equilibrium partition coefficients (5\% and 9\% error reduction), glass transition temperatures (22\% error reduction), and enthalpy of vaporization (61\% error reduction) on four datasets with atmospheric compounds. Feature analysis shows that  saturation vapor pressure and partition coefficients are governed by carbon number and oxygen-related features, whereas other phase-transition properties (e.g., enthalpy of vaporization, glass transition temperature) depend on carbon–hydrogen bond types and the presence of heteroatoms other than oxygen. This highlights the generalizability of ATMOMACCS across different datasets and properties as an interpretable molecular descriptor.

\end{abstract}

\maketitle

\section{\label{sec:int}Introduction\protect\\}
Atmospheric aerosol particles contribute to climate change by scattering and absorbing sunlight and serving as cloud condensation nuclei.\cite{IPCC2022} The presence of particles in the atmosphere also worsens air pollution and human health.\cite{Pozzer2023} Determining which compounds drive aerosol formation is an ongoing research challenge, particularly because the number of atmospheric compounds is estimated at 10$^5$–10$^6$.\cite{Goldstein2007}

The process leading to aerosol formation and growth includes atmospheric compounds originating from natural and anthropogenic emissions.\cite{Hallquist2009}  20-90 \% of the total formed sub-micron particle mass can consist of organic compounds. \cite{Chen2022,Crippa2013,Zhang2011,Jimenez2009,Zhang2007}    In the atmosphere, emissions of organic compounds undergo chemical transformations, particularly oxidation, producing a wide range of reaction products that are typically larger and chemically more complex in terms of elemental composition and functional groups.\cite{Bianchi2019} Consequently, formed oxidized compounds generally have lower volatility and a propensity to condense into the particulate phase, forming so-called secondary organic aerosols.\cite{Ehn2014, Kroll2008, Donahue2012basis}

\begin{figure*}[]
\includegraphics[width=17cm]{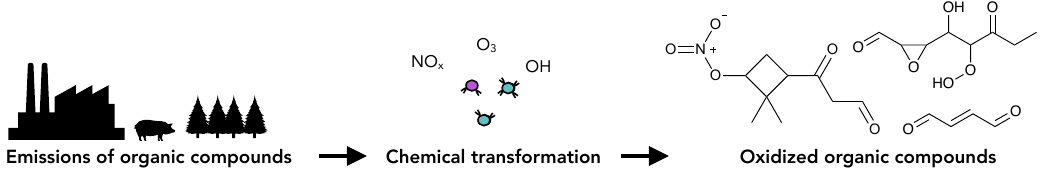}
\caption{Molecular emissions from various sources enter the atmosphere, where they undergo reactions with molecules like ozone and hydroxyl radicals. These reactions produce oxidized compounds with diverse functional groups, which are key to understanding atmospheric processes leading to particle formation. }
\label{fig:intro_fig_1}
\end{figure*}

One strategy to identify candidate atmospheric compounds that likely contribute to secondary aerosol formation is to screen their physicochemical properties related to particle formation, such as saturation vapor pressures ($P_{sat}$) and equilibrium partition coefficients.\cite{Besel2023,Besel2024,li2024,Kruger2022,Kruger2025preprint,Lumiaro2021} However, experimentally characterizing these properties is challenging. Laboratory measurements are slow and labor intensive, producing datasets with only hundreds to thousands of species (e.g., \cite{Pankow2008}),  far fewer than the estimated hundreds of thousands of atmospheric compounds.\cite{Goldstein2007}  

Computational methods provide an alternative route for high-throughput screening and property prediction. These approaches range from empirical models, which offer rapid but potentially crude estimates, to quantum chemistry calculations, which yield more precise predictions at higher computational cost.\cite{Klamt1993, Klamt1998, cosmotherm, Pankow2008, Compernolle2011}  By suggesting molecules for further experimental studies, computational methods help bridge the gap between experimental feasibility and atmospheric complexity.

Among computational approaches, group contribution methods are widely used in atmospheric science to estimate molecular properties by adding the effects of predefined structural groups. These methods have been applied to predict properties such as $P_{sat}$, enthalpies of vaporization ($\Delta H_{vap}$), refractive indices, molar volumes, densities, viscosities, and glass transition temperatures ($T_g$).\cite{Pankow2008,RamírezVerduzco2022,Li2020,Cai2017,Su2017}  A notable group contribution method is SIMPOL, which estimates $P_{sat}$ based on 30 predefined structural group terms.\cite{Pankow2008}  Other $P_{sat}$ estimation methods include EVAPORATION,\cite{Compernolle2011} Nannoolal,\cite{Nannoolal2008} Myrdal and Yalkowsky,\cite{Myrdal1997} and Tochigi.\cite{Tochigi2010} Group contribution methods, though efficient, are limited by the small datasets on which they were parametrized\cite{Pankow2008, Li2020, Compernolle2011} and show reduced accuracy against experimental benchmarks,\cite{Bannan2017, Kurten2016} partly because they ignore relative positioning of functional groups.

A new wave of models instead uses machine learning for property prediction of atmospheric compounds.\cite{Bortolussi2025, Hyttinen2022ML, Lumiaro2021, Besel2023, Elm2020, Sandstroem2025} Unlike traditional group contribution methods, which are often linear, machine learning models can capture complex, nonlinear relationships between molecular structures and properties. The success of machine learning models depends on molecular representations, or descriptors, which convert chemical structures into numerical formats that models can process.\cite{Mikulskis2019}

Molecular descriptors range from simple one-dimensional properties (e.g., molecular weight) and two-dimensional molecular fingerprint vectors to complex three-dimensional descriptors reflecting spatial atomic arrangements and interactions.\cite{Morgan1965, Rogers2010, Carhart1985, PBFingerprint2009, Landrum2022, Accelrys2011, Rupp2012, Huo2022} Dictionary-based molecular fingerprints, including MACCS,\cite{Accelrys2011} PubChem,\cite{PBFingerprint2009} and Klekota-Roth,\cite{Klekota2008} are valued for their interpretability because they encode the presence or absence of functional groups in a straightforward, human-readable way. Such interpretability is a highly-sought after trait in machine learning, providing both insights and confidence to model predictions.

Many of the available molecular descriptors were originally designed for general organic chemistry and may overlook structural characteristics of atmospheric molecules.\cite{Sandstroem2025} Atmospheric compounds often contain many oxygen- and nitrogen-rich functional groups that influence their chemistry.\cite{Bianchi2019} Omitting these features can reduce predictive accuracy and limit understanding of model predictions.

Thus, combining machine learning with group contribution approaches for atmospheric applications could improve predictive performance by leveraging the strengths of both strategies. For example, Krüger et al. \cite{Kruger2025preprint} integrated SIMPOL group contributions with graph neural networks (GNNs), which learn molecular properties from molecular graphs, and achieved more accurate $P_{sat}$ predictions. Despite this, such hybrid strategies remain largely unexplored in atmospheric chemistry. To address this need, we introduce ATMOMACCS, a molecular descriptor that combines the interpretability of the MACCS fingerprint with motifs derived from the SIMPOL group contribution method.\cite{Pankow2008} By integrating atmospheric specific motifs into a dictionary based-fingerprint, ATMOMACCS captures structural features characteristic of atmospheric compounds while retaining interpretability and computational efficiency.

\begin{figure*}[]
\includegraphics[width=8.5cm]{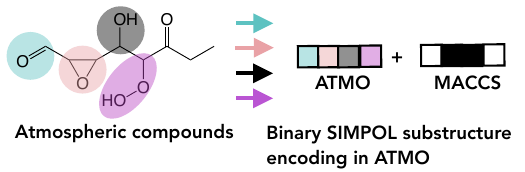}
\caption{Molecular substructures present in atmospheric compounds can be identified and incorporated into a binary molecular representation called MACCS. By extending MACCS to include additional features relevant to atmospheric chemistry, we create a new representation, ATMOMACCS.}
\label{fig:intro_fig_2}
\end{figure*}

We expect that combining MACCS with SIMPOL based features will improve molecular property predictions for atmospheric compounds. To explore this, we evaluate the predictive performance of ATMOMACCS for multiple property prediction tasks, including $P_{sat}$, equilibrium partition coefficients, $\Delta H_{vap}$, and $T_g$. A further objective is to refine the descriptor design, specifically testing different ways of encoding the atmospheric specific motifs into a machine learning ready fingerprint while retaining interpretability and computational efficiency.

This paper is organized as follows: Section \ref{sec:met} describes the design and testing of ATMOMACCS together with the machine learning methodology. Section \ref{sec:res} presents the predictive performance of ATMOMACCS, compares ATMOMACCS to traditional molecular descriptors, and demonstrates model interpretability through feature importance analysis. Section \ref{sec:dis}  highlights the strengths and limitations of ATMOMACCS and suggests directions for further development, and Section and \ref{sec:con} provides conclusions.

\section{\label{sec:met} Methods\protect\\}

\subsection{\label{sec:datasets}Datasets}

We developed ATMOMACCS using atmospheric molecular datasets to ensure its suitability for machine learning applications in atmospheric science. We compiled four datasets of atmospheric compounds and their properties from the literature (see Table \ref{tab:datasets}). These datasets focus exclusively on organic compounds with experimentally measured or computationally predicted properties. The \textit{Wang} dataset, compiled by Wang et al.,\cite{Wang2017} contains 3414 atmospheric compounds generated using the Master Chemical Mechanism code\cite{MCM} by simulating the oxidation of 143 volatile organic compounds, including methane and 142 non-methane species. In the \textit{Wang} dataset, each molecule is associated with three computed properties: $P_{sat}$, water-to-gas equilibrium partition coefficient ($K_{W/G}$), and water-insoluble organic matter to gas equilibrium partition coefficient ($K_{WIOM/G}$). These properties have been computed with the quantum chemistry based method COSMOtherm \cite{Klamt1998, Klamt1993} at 288.15 K (see ref. \cite{Wang2017} for computational details). Similarly, the \textit{GeckoQ} dataset\cite{Besel2023} contains 31637 oxidized organic molecules with their $P_{sat}$ predicted by COSMOtherm at 298.15 K. \textit{GeckoQ} is a subset of a larger 167k molecule dataset generated by the Gecko-A code (Generator for Explicit Chemistry and Kinetics of Organics in the Atmosphere, https://geckoa.lisa.u-pec.fr/index.php),\cite{IsaacmanVanwertz2020, Besel2023} which simulates atmospheric oxidation. The \textit{GeckoQ} dataset was generated starting from three volatile organic compound precursors: $\alpha$-pinene, decane, and toluene. The third dataset, by Ferraz-Caetano et al,\cite{FerrazCaetano2024} includes 2410 molecules, including 223 volatile organic compounds, with experimentally measured $\Delta H_{vap}$.\cite{Acree2010, Gharagheizi2013,Marlowe1995} Finally, \textit{Li et al.}\cite{Li2020} curated a dataset for computing $T_g$ with 2718 atmospherically relevant compounds. $T_g$ relates to molecular viscosity and thereby impacts aerosol properties. After filtering out missing values, the dataset was reduced to 2216 compounds for our purposes. In the $Li$ dataset, the $T_g$ values are a mixture of computational predictions, experimental measurements, and estimates derived from melting points.\cite{Li2020} 

\begin{table*}
\caption{\label{tab:datasets}The four molecular datasets used for benchmarking ATMOMACCS. Listed are the dataset names used in this paper, number of compounds, associated target properties, relevant temperatures, whether the dataset target was collected with computational or experimental methods, as well as reference. Acronyms:  P$_{sat}$ - saturation vapor pressure;  $K_{W/G}$ - water-gasphase equilibrium partition coefficient; K $_{WIOM/G}$ - water insoluble organic matter - gasphase equilibrium partition coefficient; $\Delta H_{vap}$ - enthalpy of vaporization; T$_g$ - glass transition temperature.}
\begin{ruledtabular}
\begin{tabular}{cccccccc}
    
 Dataset &  Size & Elements present &  Assoc. target property & Temp. [K] & Comp. data & Exp. data & Ref.  \\ \hline 
    
   GeckoQ & 31637 & C, H, N, O & P$_{sat}$ & 298.15 & Yes & No & \citep{Besel2023} \\    
    
   Wang & 3314 & C, H,  N, O, S, Cl, Br &   P$_{sat}$,  K$_{W/G}$, K$_{WIOM/G}$ & 288.15 & Yes & No & \citep{Wang2017}   \\
    
    Ferraz-Caetano & 2410 & C, H, N, O, F, P, S, Cl, Br, I & $\Delta H_{vap}$ & 298.15  & No & Yes & \citep{FerrazCaetano2024} \\
    
    Li &  2216 & C, H, N, O, Na, S, Cl & T$_{g}$  &  N/A & Yes  & Yes & \citep{Li2020}\\
\end{tabular}
\end{ruledtabular}
\end{table*}

\begin{figure*}[]
\includegraphics[width=17cm]{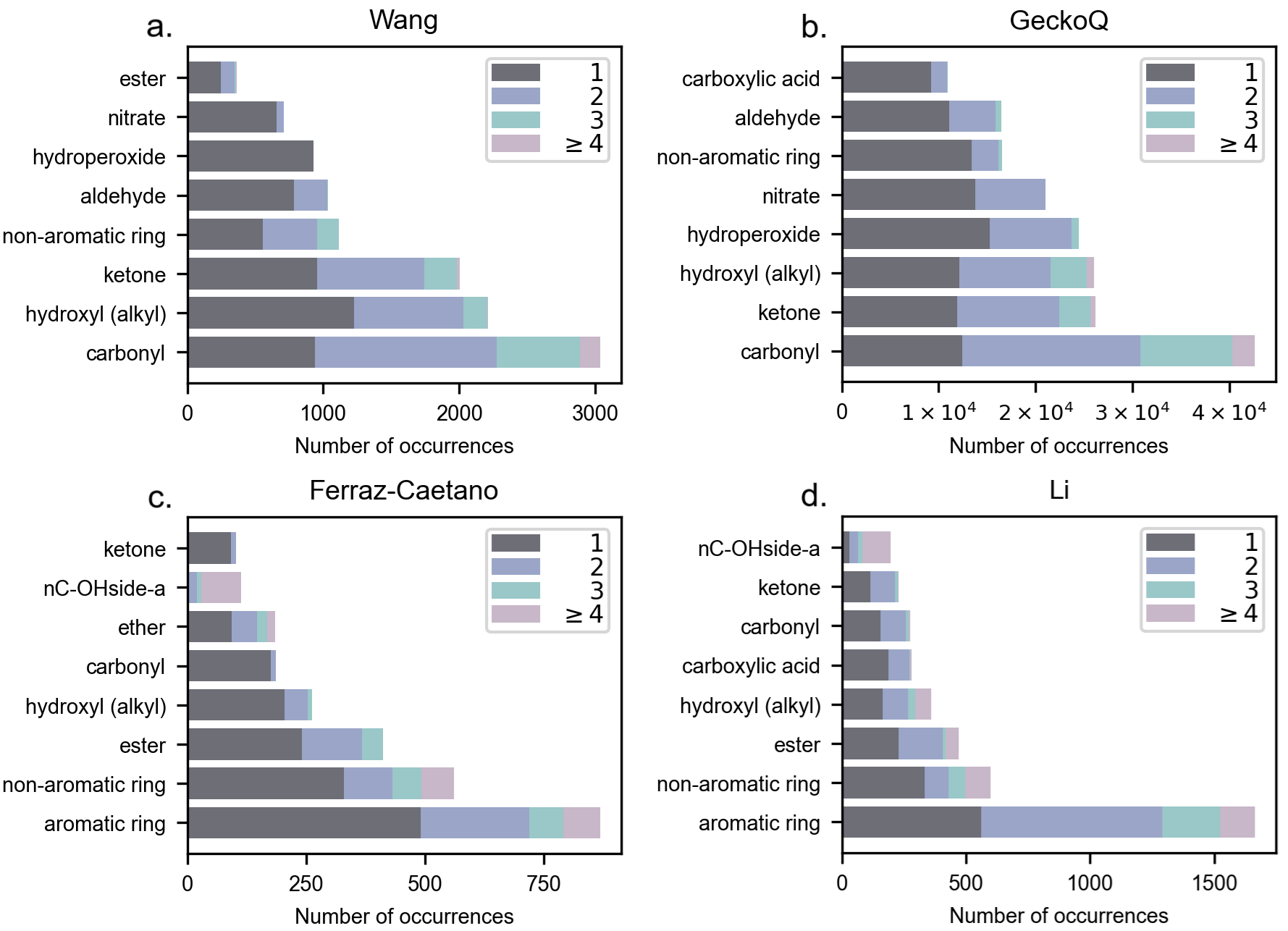}
\caption{Functional fragment counts (SIMPOL) for (a) \textit{Wang},\cite{Wang2017} (b) \textit{GeckoQ},\cite{Besel2023} (c) \textit{Ferraz-Caetano},\cite{FerrazCaetano2024} and (d) \textit{Li}\cite{Li2020} datasets. Carbon number, oxygen number and carbon types are excluded. The colors represent the number of occurrences within the same molecule. In the datasets, most fragments appear once or twice per molecule on average.}
\label{fig:fun_groups}
\end{figure*}

\begin{figure*}[]
    \includegraphics[width=17cm]{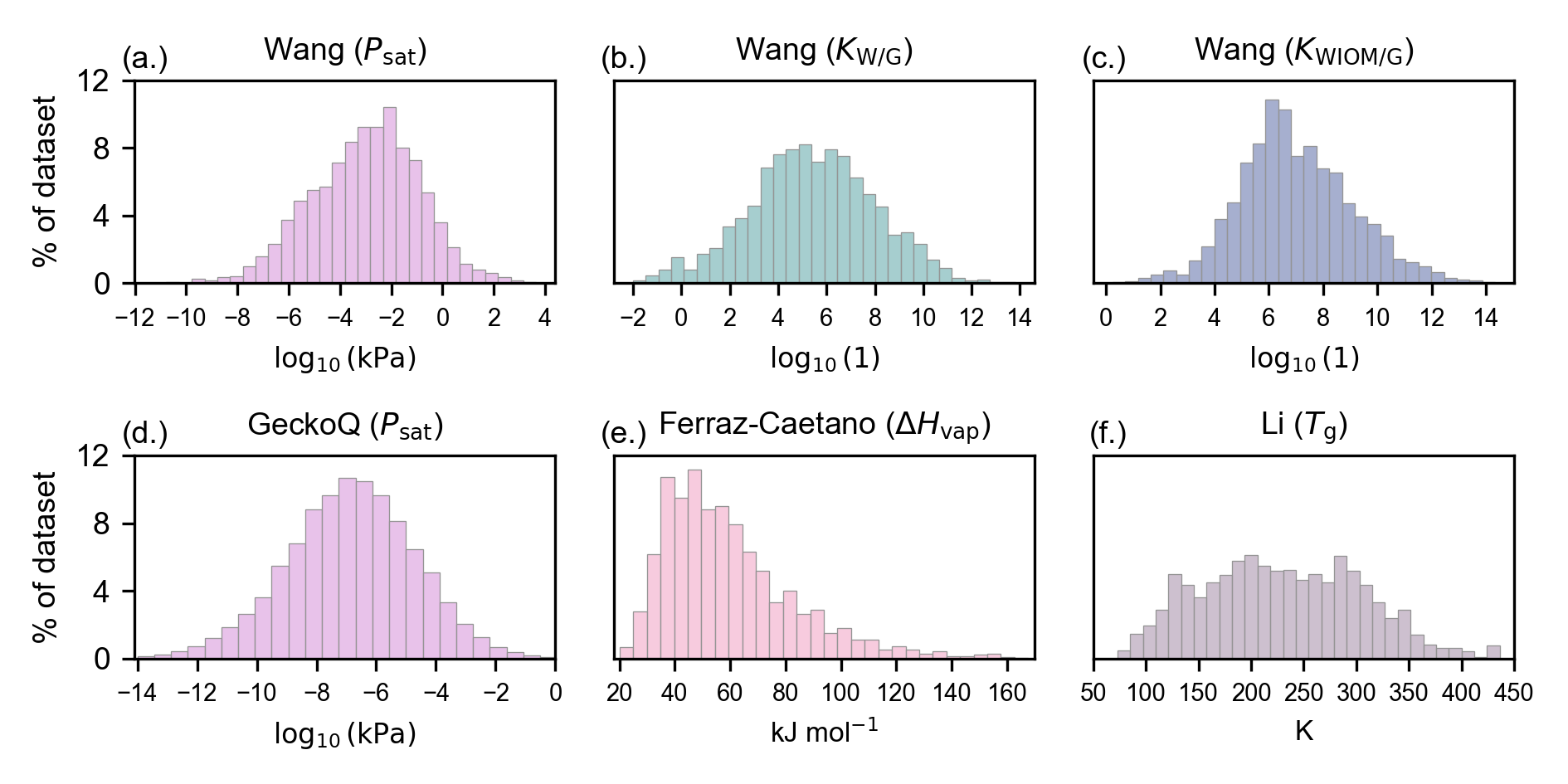}
    \caption{Distribution of target values in the datasets: (a) saturation vapor pressure ($P_{sat}$) from the \textit{Wang} dataset \cite{Wang2017}, (b) water–gas equilibrium partition coefficient ($K_{W/G}$) from the \textit{Wang} dataset, (c) water-insoluble organic matter–gas equilibrium partition coefficient ($K_{WIOM/G}$) from the \textit{Wang} dataset, (d) saturation vapor pressure ($P_{sat}$) from the \textit{GeckoQ} dataset \cite{Besel2023}, (e) enthalpy of vaporization ($\Delta H_{vap}$) from the \textit{Ferraz-Caetano} dataset, \cite{FerrazCaetano2024} and (f) glass transition temperature ($T_{g}$) from the dataset of Li et al. \cite{Li2020}
}
    \label{fig:test1}
\end{figure*}

In Table \ref{tab:datasets}, we present the elemental composition of the molecular datasets. The datasets consist of organic atmospheric compounds that are primarily composed of carbon and oxygen, with varying amounts of oxygen reflecting different degrees of oxidation. For instance, the hydrocarbons in \textit{Ferraz-Caetano} contain a minimal number of oxygen atoms,  whereas \textit{GeckoQ} includes highly oxygenated compounds. Elements such as nitrogen, sulfur, and chlorine appear in smaller amounts in \textit{Wang}, \textit{Li} and \textit{Ferraz-Caetano}, whereas \textit{GeckoQ} contains exclusively carbon, oxygen, and nitrogen atoms. Functional groups analysis (Fig. \ref{fig:fun_groups}) reveals that the most common oxygen- and nitrogen-containing functional groups in the different datasets are ester (including nitroester), hydroxyl, ketone, and carbonyl groups. Figure \ref{fig:size_dist_appendix} in Appendix A shows the size distribution of compounds across the four datasets, with sizes mostly ranging from two to 27 non-hydrogen atoms with an average of 11 to 18 non-hydrogen atoms. A few outliers  in \textit{Ferraz-Caetano}'s and \textit{Li}'s datasets reach sizes up to 82 non-hydrogen molecules, which we kept in the dataset. Figure \ref{fig:test1} shows the distributions of the molecular properties for the four datasets, which we later use as modeling targets. Note that we use a log scale for the pressure (\textit{kPa}) and equilibrium coefficients in the figures and for model development.

\subsection{\label{sec:met_simpol} The SIMPOL group contribution method} 
ATMOMACCS incorporates atmospheric chemistry domain knowledge through the SIMPOL group contribution method.\cite{Pankow2008} As mentioned in the introduction, SIMPOL is a parametrized model for estimating $P_{sat}$ based on a set of molecular substructures. SIMPOL considers functional groups such as aldehydes, ketones, esters, carboxylic acids, nitrates, and peroxides, for a total of 30 substructures in the original publication. In SIMPOL, $P_{sat}$ is computed from the number of occurrences $n_i$ of each substructure $i$ and its associated contribution $\Delta \log_{10}(P_i)$, as 
\begin{eqnarray}
\log_{10}(P_{sat}) = \sum_{i} n_i \cdot \Delta \log_{10}(P_i),
\label{eq:simpol}
\end{eqnarray}
where the $\Delta \log_{10}(P_i)$ terms have been fitted to experimental data (and include a temperature dependence).\cite{Pankow2008} Table \ref{tab:atmo_groups} lists the substructure groups from the SIMPOL implementation in the APRL Substructure Search Program (aprl-ssp)\cite{Ruggeri2016, Takahama2017}   which were used in our descriptor development.

\begin{table*}
\caption{\label{tab:atmo_groups} Substructure groups taken from the SIMPOL implementation of the APRL Substructure Search Program (aprl-ssp).\cite{Ruggeri2016, Takahama2017} The ATMO keys in ATMOMACCS are based on this list but have removed certain redundant information for the machine learning model, see footnotes. We have enumerated the substructures for reference. Keys marked with a dagger ($^{\dagger}$) were not part of the original SIMPOL publication\cite{Pankow2008} but were included in the APRL Substructure Search Program, with the exception of the oxygen count, which we added.}

\begin{ruledtabular}
\begin{tabular}{r l r l r l}
1  & Zeroeth group\footnotemark[1] & 15 & Nitro & 29 & Ether (alicyclic) \\
2  & Amine, primary & 16 & Aromatic hydroxyl & 30 & Amine, aromatic \\
3  & Amine, secondary & 17 & Hydroperoxide & 31 & Nitroester \\
4  & Amine, tertiary & 18 & Amide & 32 & C=C-C=O in non-aromatic ring \\
5  & Alkane CH$^{\dagger}$ & 19 & Nitrate & 33 & C=C (non-aromatic) \\
6  & Alkene CH$^{\dagger}$ & 20 & Organosulfate & 34 & Number of carbon atoms in side chain(s) attached to an amide nitrogen \\
7  & Aromatic CH$^{\dagger}$ & 21 & Ketone & 35 & Carbon number on the acid-side of amide (asa)\footnotemark[3] \\
8  & Carbonyl & 22 & Aldehyde & 36 & Carbonylperoxynitrate \\
9  & Hydroxyl (alkyl) & 23 & Amide, primary & 37 & Nitrophenol \\
10 & Carboxylic acid & 24 & Amide, secondary & 38 & Number of carbons\footnotemark[4] \\
11 & All esters$^{\dagger}$ & 25 & Amide, tertiary & 39 & Aromatic ring \\
12 & Ester\footnotemark[2] & 26 & Carbonylperoxyacid & 40 & Non-aromatic ring \\
13 & Ether & 27 & Peroxy nitrate & 41 & Number of oxygen atoms$^{\dagger}$\footnotemark[4] \\
14 & Peroxide & 28 & Ether, aromatic & & \\
\end{tabular}
\end{ruledtabular}

\footnotetext[1]{Intercept term of SIMPOL. ATMO does not include this key.}
\footnotetext[2]{Excluded from ATMO due to redundancy with ‘all esters’ and ‘nitroester’ motifs.}
\footnotetext[3]{Excluded from ATMO for practicality and relevance. Present in Ferraz-Caetano and Li datasets, but not Wang or GeckoQ.}
\footnotetext[4]{Only in ATMO versions 3 to 5.}

\end{table*}

\subsection{\label{sec:met_maccs}MACCS }
ATMOMACCS combines the SIMPOL domain knowledge with the MACCS structural keys, a set of 166 binary features indicating the presence or absence of specific functional groups, elements, and their relative positions within a molecular structure. Some MACCS keys also detect isotopes or multiple fragments (relevant for, e.g., salts). A full description of all 166 keys is provided in the MACCS whitepaper\cite{Accelrys2011}. Originally developed in the 1990s by MDL Information Systems (now BIOVIA), the MACCS fingerprint has been widely implemented in toolkits such as CDK, OpenBabel, \cite{Oboyle2011} and RDKit.\cite{Landrum2022} Our work uses the RDKit implementation, which includes a 167th dummy key that we retain for consistency. MACCS provides a compact structural representation but was not developed for oxidized atmospheric organic compounds that often contain a diverse array of oxygen-bearing functional groups (Figure \ref{fig:fun_groups}).

\subsection{\label{sec:met_atmomaccs}Development of ATMOMACCS}
To create ATMOMACCS, we extend the MACCS fingerprint with a  new set of  features which captures the molecular structure of atmospheric compounds (ATMO, see Table \ref{tab:atmo_groups}), producing a combined representation. Throughout this paper, the terms key and feature are used interchangeably.  We have developed ATMOMACCS in two formats: a binary fingerprint and a numerical representation. The binary fingerprint encodes yes-or-no answers to questions about molecular features, while the numerical representation records the absolute counts of each motif. Each format has its advantages: the binary version efficiently captures molecular motifs for large datasets, whereas the numerical version provides more detailed information at a higher computational cost.  Here, ATMOMACCS is evaluated in five versions, each designed for specific applications and insight. The details of each version is reported in Table \ref{tab:atmoversions}.

\begin{table*}
\caption{\label{tab:atmoversions}The five ATMOMACCS versions evaluated in this work. Listed are the version name, total number of keys, encoding scheme, and key differences between versions.}
\begin{ruledtabular}
\begin{tabular}{cccc}
Version & Keys & Encoding & Key differences \\ \hline
V1 & 202 & Binary & Presence/absence of SIMPOL groups \\
V2 & 274 & Binary & Presence/absence of SIMPOL groups in 0, 1, 2, or $>$2 instances \\
V3 & 280 & Binary & V2 plus binary encoding of carbon atom count (up to 63) \\
V4 & 286 & Binary & V3 plus binary encoding of oxygen atom count (up to 63) \\
V5 & 204 & Integer & Counts of all MACCS and ATMO keys (replaces binary encoding) \\
\end{tabular}
\end{ruledtabular}
\end{table*}

\begin{figure*}[]
\includegraphics[width=17cm]{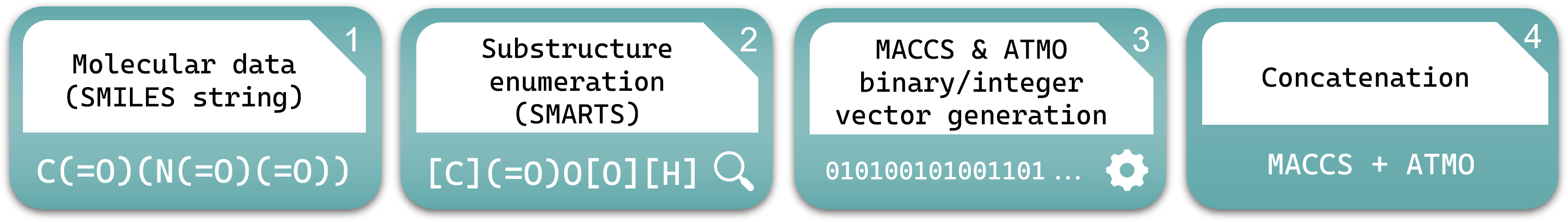}
\caption{The construction of ATMOMACCS follows a four step process. First, the two dimensional molecular structure is read from the molecular SMILES (Simplified Molecular Input Line Entry System) string. Next, the appearance of ATMO features is counted based on the specified molecular structure. These counts are then converted into a binary representation, with the specific encoding scheme varying across different ATMOMACCS versions. Finally, the MACCS fingerprint is concatenated with the ATMO keys to produce the complete ATMOMACCS molecular descriptor.}
\label{fig:AM_wflow}
\end{figure*}

We have developed a custom code that identifies and counts appearances of ATMO groups based on the APRL Substructure Search Program (aprl-ssp).\cite{Ruggeri2016, Takahama2017}   Our ATMOMACCS implementation uses Python version 3.12.5 and depends on RDKit version 2023.09.1.
The practical construction of ATMOMACCS follows the workflow illustrated in Figure \ref{fig:AM_wflow}. First, we read molecular structures from their SMILES (Simplified Molecular Input Line Entry System) representations. Next, we scan each structure for SIMPOL motifs using SMARTS (SMILES arbitrary target specification) patterns. We then generate MACCS and ATMO fingerprints in binary format for versions 1–4 and in integer format for version 5. Finally, we concatenate the ATMO and MACCS features to form the combined ATMOMACCS fingerprint. The implementation closely follows the original MACCS fingerprint python implementation in RDKit (rdkit.Chem.MACCSKeys), ensuring easy use and transferability.

\subsection{\label{sec:met_ml} Machine learning model training and evaluation}
We evaluate ATMOMACCS by testing model performance in different property prediction tasks.  We adopt the kernel ridge regression (KRR) algorithm for our machine learning model, building on our previous work with the \textit{Wang} dataset\cite{Lumiaro2021}. KRR is trained using a dataset comprising input features and corresponding target values. In our context, the input features are molecular descriptors, while the target values represent molecular properties. The KRR method extends ridge regression, which applies a penalty term to the least-squares fit to prevent overfitting. By incorporating a non-linear kernel, KRR effectively models non-linear relationships. However, the training process scales roughly as \( O(n^3) \), where \( n \) represents the number of training inputs. The matrix inversion required to compute regression coefficients makes training KRR on large datasets time- and memory-consuming.

Several hyperparameters influence the performance of KRR. In this work, we use a Gaussian kernel and optimize the regularization (penalty) parameter \( \lambda \) along with the Gaussian kernel specific parameter \( \gamma \) for each dataset and descriptor combination using grid search.

We evaluate ATMOMACCS and other descriptors by comparing the mean absolute error (MAE) of the KRR model on the test set. The test set MAE measures the magnitude of the error for unseen data in units of the predicted quantity. Thus, this error metric produces a physically meaningful true error estimate for property prediction.

We implement KRR using scikit-learn\cite{Pedregosa2011} and employ random train-test splits. For all datasets, we reserve a fixed test set of 12 \% of the dataset while varying the size of the training set between 15 \% and 88 \% of the dataset in six linear increments. With this train-test split procedure, we examine the effectiveness of the descriptors by analyzing the learning curves obtained from training the KRR model at varying training set sizes. These curves quantitatively indicate improvements in the model as more data is allocated for training. We compare the performance of ATMOMACCS with other descriptors by assessing these learning curves, particularly focusing on the MAE metric for the largest training set size. We average these results across ten random samplings of the training and test set data to mitigate the effects of random splits.

\subsection{\label{sec:met_rep}Reference Molecular Representations}
In our benchmark of ATMOMACCS, we compare models trained on other descriptors. In particular, we include the topological fingerprint which has previously shown good performance for $P_{sat}$ and equilibrium partition coefficient predictions. \cite{Lumiaro2021, Besel2023, Besel2024}

Similarly to MACCS and ATMOMACCS, the topological fingerprint\cite{Landrum2022} is a two-dimensional molecular descriptor.  However, the features of the topological fingerprint are determined by enumerating possible paths in the molecular structure, which are then hashed into a binary representation. Although the path-bit correspondence can be deduced, the absence of a one-to-one mapping complicates its chemical interpretability, because the paths do not directly align with chemically meaningful substructures, such as functional groups.

The performance of the topological fingerprint can be fine-tuned by optimizing its hyperparameters, including fingerprint length, bits per hash, and minimum and maximum path lengths. Previously, we found that the topological fingerprint was relatively insensitive to hyperparameter choices when it was used to train a KRR model on the \textit{Wang} dataset.\cite{Wang2017} Here we have optimized the fingerprint length, bits per hash, as well as minimum and maximum path lengths, using grid search.

In our benchmark of ATMOMACCS, we compare with models trained on MACCS and ATMO features alone to identify which combination of features results in the most accurate model. In addition, this comparison can validate our approach, which combines specific features related to atmospheric chemistry (ATMO) with more general chemistry features (MACCS). When comparing to the standalone ATMO features, we chose the version 5 set of features (see Table \ref{tab:atmoversions}).

\subsection{\label{sec:met_shap}Shapley Additive Explanations Analysis}
An interpretable molecular representation with chemically meaningful features enables the use of modern feature importance analysis tools to provide chemical insight from machine learning models. We employ SHAP (SHapley Additive exPlanations)\citep{Lundberg2017, Lundberg2019arxiv} value analysis to assess the contributions of these molecular fingerprint features to the predictions made by the KRR model.\citep{Lundberg2017, Lundberg2019arxiv} SHAP values are calculated by varying feature values and observing changes in model predictions. We note that SHAP values can be either positive or negative, reflecting their directional effect on the predicted property. In this work, however, we focus exclusively on the magnitude of SHAP values when presenting and discussing feature importance in the text. References hereafter to \textit{high} or \textit{low} SHAP values therefore pertain only to their magnitude. Features with minimal impact on the output will have low SHAP values. Conversely, features with a large effect on predictions will have high SHAP values, highlighting their important role in the model's decision-making process. With the SHAP analysis we obtain feature importance values for all ATMOMACCS features. We implemented the SHAP analysis using the SHAP library in Python.\citep{Lundberg2017, Lundberg2019arxiv}  SHAP allows us to interpret the KRR model predictions and find correlations between molecular features and properties.

\section{\label{sec:res} Results\protect\\}
To assess the utility of ATMOMACCS, we first benchmark its different versions (Table \ref{tab:atmoversions}) on a series of property prediction tasks. We also compare its performance with other molecular descriptors using the same evaluation scheme. The tasks include predictions of $P_{sat}$, $K_{W/G}$, $K_{WIOM/G}$, $\Delta H_{vap}$, and $T_{g}$. Finally, we apply SHAP analysis (see Section \ref{sec:met}) to identify the most influential molecular features and gain a deeper understanding of ATMOMACCS performance. Our model results are also summarized in Table \ref{tab:mae-comparison} in Appendix A.

\subsection{\label{sec:res_psat}Saturation vapor pressures}

\begin{figure*}[]
\includegraphics[width=17cm]{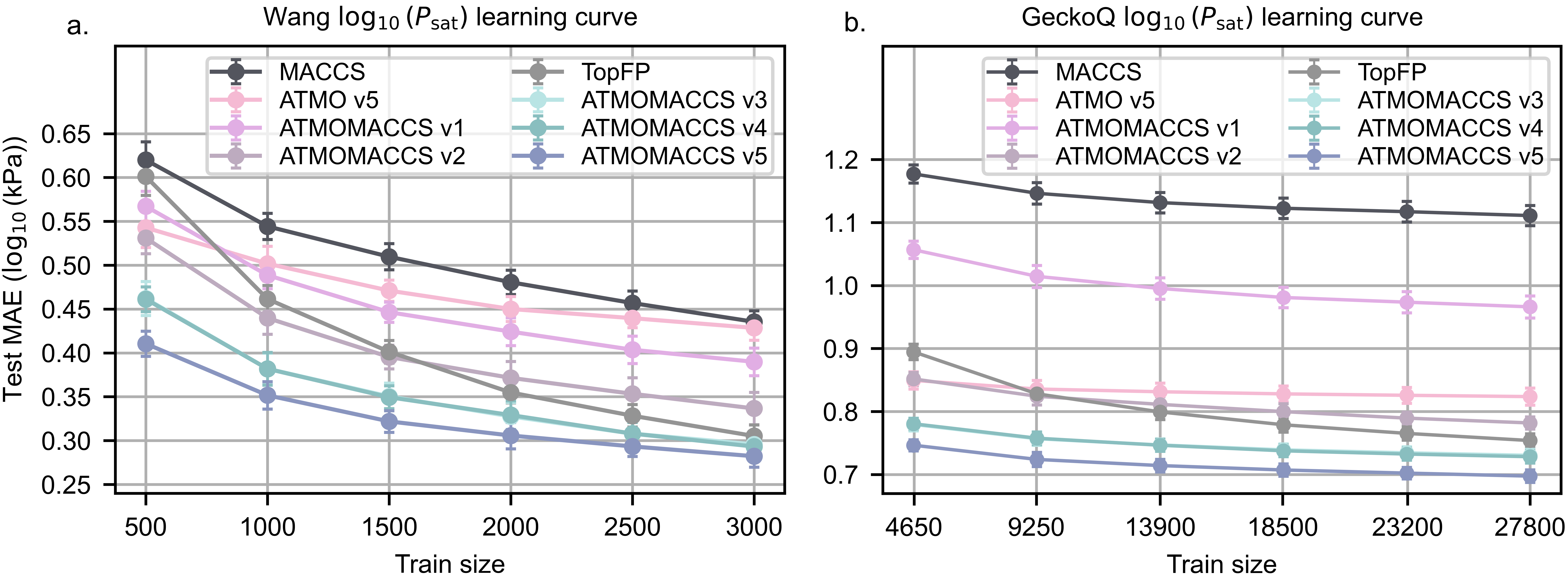}
\caption{(a) The learning curves of the machine learning (kernel ridge regression, KRR) prediction model for $P_{sat}$ prediction on the \textit{Wang} dataset\cite{Wang2017} using different molecular fingerprints. (b) The learning curves of the machine learning (KRR) prediction model for $P_{sat}$ prediction on the \textit{GeckoQ} dataset\cite{Besel2023} using different molecular fingerprints. The vertical axis shows the mean absolute error (MAE) of model predictions on the test set. The error bars correspond to the standard deviation across 10 runs with different random seeds.}
\label{fig:lc_psat}
\end{figure*}

In Figure \ref{fig:lc_psat}, we show the learning curves of our KRR $P_{sat}$ predictions.  In Figure \ref{fig:lc_psat}a and \ref{fig:lc_psat}b we present models that have been trained on the \textit{Wang} and \textit{GeckoQ} datasets, respectively. The figure shows accuracy in the form of MAE  when the model is trained using ATMOMACCS, ATMO, MACCS and the topological fingerprint for different training set sizes. For all descriptors, we observe learning when the training set size increases, as seen by the MAE decreasing.  

% ATMOMACCS VERSIONS
First, we compare the relative performance of our KRR models with the different ATMOMACCS versions (see Table \ref{tab:atmoversions})  for $P_{sat}$ predictions at the largest training set size in Figure \ref{fig:lc_psat}. These results are also summarized in Table \ref{tab:mae-comparison}. Performance trends are the same for both the \textit{Wang} and \textit{GeckoQ} datasets, with increased performance (lower MAE) on the test set for each successive ATMOMACCS version. In Panel a of the \textit{Wang} dataset, a notable improvement of 0.05 $\log_{10}$(kPa) is observed between versions 1 and 2. This enhancement arises from considering not only the presence of ATMO features (Table \ref{tab:atmo_groups}), but also their frequency—whether they appear once, twice, or multiple times. Including the carbon atom count in version 3 further reduces the error. However, adding the oxygen number (version 3 to 4) gives only a marginal improvement. The integer encoded version 5 further reduces the error, for a final MAE of 0.28 log units. 

For \textit{GeckoQ}, the MAE reduction is the largest between versions 1 and 2 (0.18 log units). Incorporating the carbon number further reduces the MAE. However, similar to the Wang dataset, adding the oxygen number (version 4) yields only a marginal gain of less than 0.01 log units. Notably, version 5 reduces the MAE to 0.70 logarithmic units. 

% ATMOMACCS AND MACCS 
We now compare ATMOMACCS to the standalone MACCS descriptor in Figure \ref{fig:lc_psat}. For both the \textit{Wang} and \textit{GeckoQ} datasets, MACCS based models have the highest MAE (Table \ref{tab:mae-comparison}), while ATMOMACCS consistently achieves lower errors in each successive version. This trend is observed for both datasets, with larger improvements for the \textit{GeckoQ} dataset. Across both datasets, ATMOMACCS consistently achieves lower MAEs than the MACCS-based model, reflecting the contribution of additional ATMO substructure features.

To test whether the ATMO features alone could provide comparable performance, we next compare ATMOMACCS to a standalone form of ATMO. In Figure \ref{fig:lc_psat}, we observe that ATMO performs similarly to MACCS alone for the \textit{Wang} dataset.  Meanwhile, for \textit{GeckoQ}, the ATMO features performs appreciably better than both MACCS and ATMOMACCS version 1, yet version 5 has an 0.12 lower MAE. For both datasets, the lowest MAEs are observed for the combination of MACCS and ATMO features in ATMOMACCS version 5.

\begin{figure*}[!htbp]
\includegraphics[width=17cm]{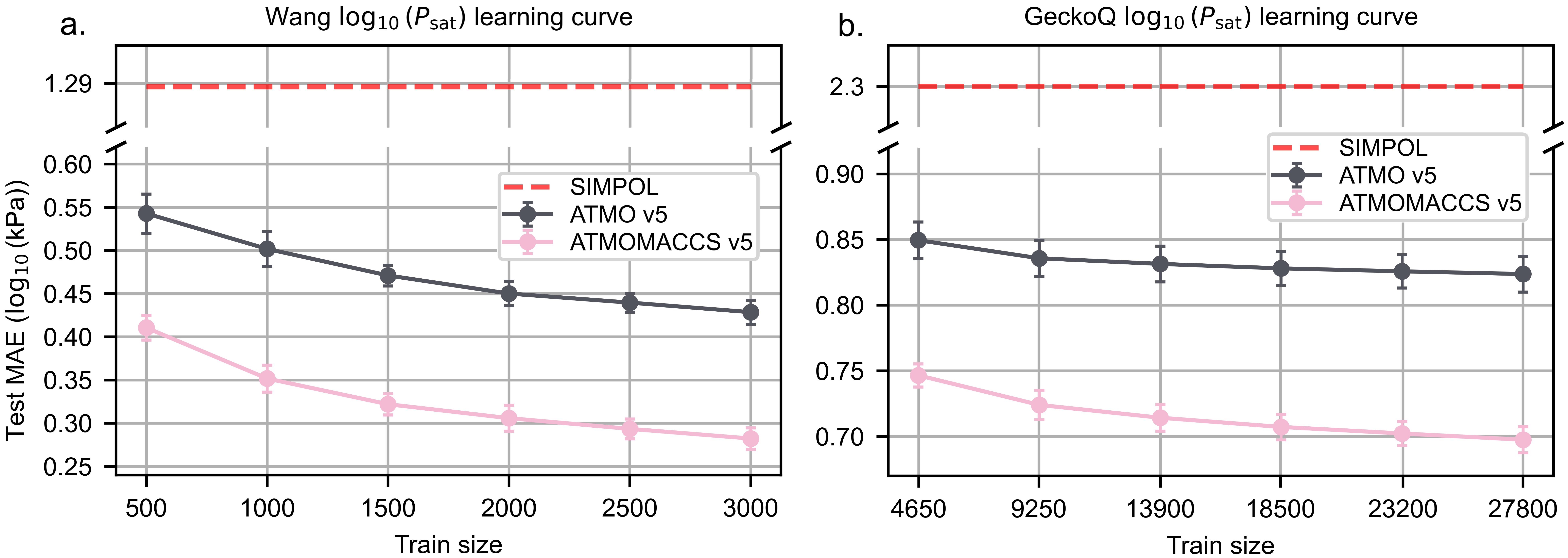}
\caption{SIMPOL\cite{Pankow2008} mean absolute error (MAE) compared to the learning curves of the machine learning (kernel ridge regression, KRR) prediction model for $P_{sat}$ prediction on the (a) \textit{Wang}\cite{Wang2017} and (b) \textit{GeckoQ} \cite{Besel2023} datasets using best performing ATMO and ATMOMACCS versions. The vertical axis shows the mean absolute error (MAE) of model predictions on the test set. The error bars correspond to the standard deviation across 10 runs with different random seeds.}
\label{fig:lc_psat_w_simpol}
\end{figure*}

Figure \ref{fig:lc_psat_w_simpol} compares the performance of SIMPOL with KRR models trained on ATMOMACCS and ATMO (both version 5) for the \textit{Wang} and \textit{GeckoQ} datasets. The ATMOMACCS-based KRR model reduces prediction errors by more than a factor of two relative to SIMPOL for both datasets. Even the KRR model using only ATMO features achieves a lower MAE than SIMPOL.

% ATMOMACCS AND TOPOLOGICAL  ()
In Figure \ref{fig:lc_psat} we compare ATMOMACCS to the topological fingerprint, which has been shown to be among the best molecular descriptors for $P_{sat}$ predictions of atmospheric compounds.\cite{Lumiaro2021}  We observe that ATMOMACCS versions 3 to 5 outperform the topological fingerprint for both datasets. In particular, ATMOMACCS version 5 achieves MAEs of 0.28 and 0.70 log units for the \textit{Wang} and \textit{GeckoQ} datasets, respectively, compared to 0.31 and 0.75 log units for the topological fingerprint. 

\subsection{\label{sec:res_parcoeff} Equilibrium partition coefficients}
\begin{figure*}[!htbp]
\includegraphics[width=17cm]{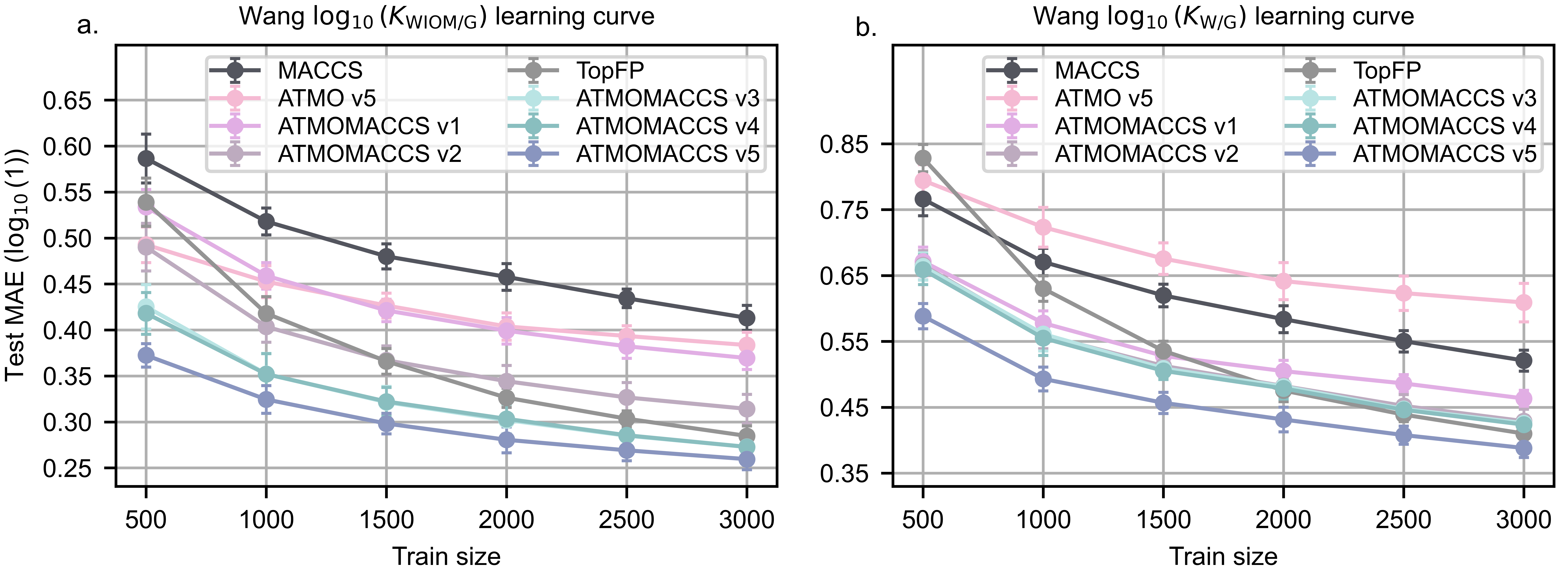}
\caption{The learning curves of the machine learning (kernel ridge regression, KRR) prediction model for equilibrium partition coefficient prediction on the \textit{Wang }dataset \cite{Wang2017} using different molecular fingerprints for (a) the water-gasphase equilibrium partition coefficient $\log_{10}(K_{W/G})$ and (b) the water insoluble matter-gasphase equilibrium partition coefficient, $\log_{10}(K_{WIOM/G})$. The vertical axis shows the mean absolute error (MAE) of model predictions on the test set. The error bars correspond to the standard deviation across 10 runs with different random seeds.}
\label{fig:lc_kwq}
\end{figure*}

We have evaluated the performance of the descriptors for predicting equilibrium partition coefficients using the \textit{Wang} dataset in Figure \ref{fig:lc_kwq}. For $\log_{10}K_{WIOM/G}$, the descriptor ranking matches that observed for $P_{sat}$ prediction (Panel a). The largest gains occur between versions 1 and 2 and between versions 2 and 3, corresponding to the addition of multiple functional group counts and carbon number, respectively. Adding oxygen number in version 4 yields only a marginal improvement. For $ \log_{10}K_{W/G}$, the trend differs (Panel b). The largest improvement again appears between versions 1 and 2, but subsequent versions 3 and 4 provide only slight gains. Version 5, however, yields an additional improvement of 0.04 log units.

When comparing descriptor sets, all ATMOMACCS versions outperform the original MACCS fingerprint. For $ \log_{10}(K_{WIOM/G})$, ATMO performs better than MACCS while the opposite is true for $ \log_{10}(K_{W/G})$. In both cases, standalone ATMO performs worse than ATMOMACCS by a substantial margin. For example, for $\log_{10}(K_{W/G})$ the learning curve plateaus early, trailing other descriptors by 0.09–0.22 log units. Finally, later ATMOMACCS versions also surpass the topological fingerprint: version 5 for $\log_{10}(K_{WIOM/G})$, and versions 4 and 5 for $ \log_{10}(K_{W/G})$, with the latter showing only a marginal advantage (0.02 log units).

\subsection{\label{sec:res_hvap} Vaporization enthalpies}
\begin{figure*}[!htbp]
\includegraphics[width=17cm]{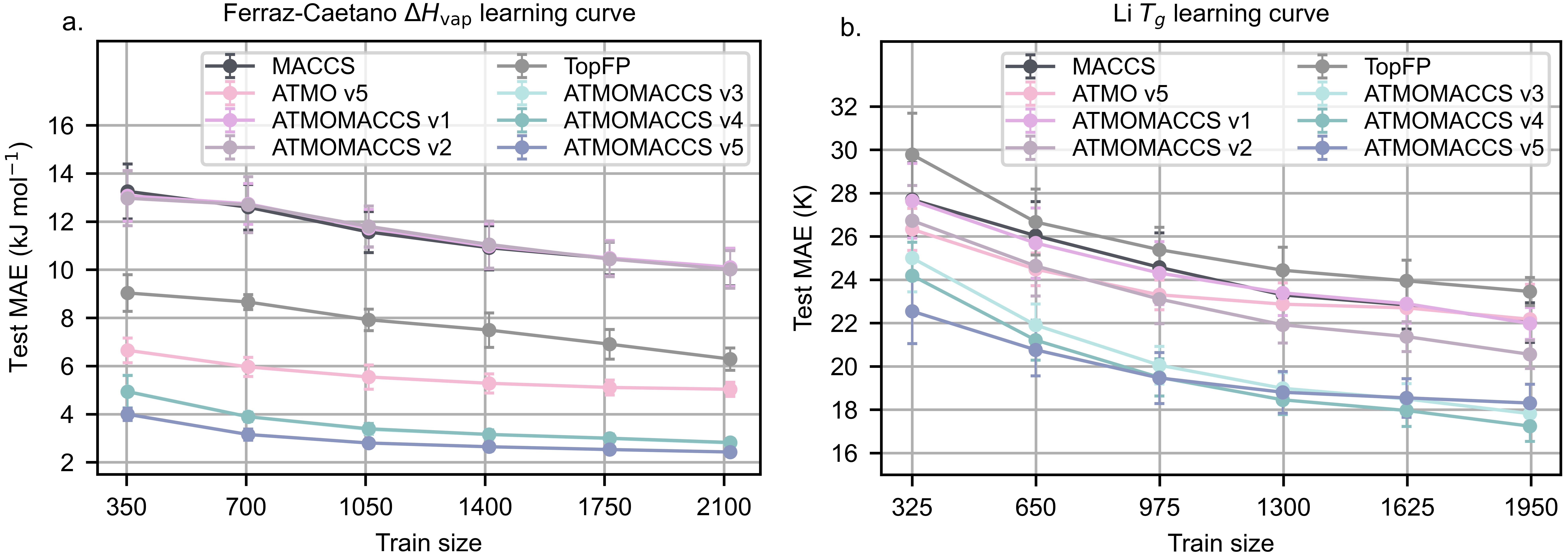}
\caption{The learning curves of the machine learning (kernel ridge regression, KRR) prediction models  using different molecular fingerprints for (a) the enthalpy of vaporization, $\Delta H_{vap}$, with the \textit{Ferraz-Caetano} et al. dataset\cite{FerrazCaetano2024} and (b) the glass transition temperature, $T_g$. and the \textit{Li} et al. dataset. The vertical axis shows the mean absolute error (MAE) of model predictions on the test set. The error bars correspond to the standard deviation across 10 runs with different random seeds.}
\label{fig:lc_dvaptg}
\end{figure*}

In Figure \ref{fig:lc_dvaptg}, we assess ATMOMACCS for predicting $\Delta H_{vap}$. Among ATMOMACCS versions, performance improves steadily from version 1 to 5 (Figure \ref{fig:lc_dvaptg}a, MAE reduced from 10.10 to 2.43 kJ mol$^{-1}$ ). Versions 1 and 2 have similar MAE, versions 3 and 4 are identical with lower MAE, and version 5 achieves the lowest MAE of all ATMOMACCS versions.

Compared to other descriptors, ATMOMACCS versions 1–2 exhibit performance nearly identical to MACCS, with MAE values of 10.02–10.10 kJ mol$^{-1}$ and differences within the error bars. The topological fingerprint yields a 3.81 kJ mol$^{-1}$ lower MAE than ATMOMACCS 1–2, but a 3.86 kJ mol$^{-1}$ higher value than ATMOMACCS 5. The standalone ATMO descriptor also performs well for this property, with an MAE of 5.03 kJ mol$^{-1}$, surpassed only by ATMOMACCS versions 3–5.

\subsection{\label{sec:res_tg} Glass transition temperature}
We next examine $T_g$ predictions (Figure \ref{fig:lc_dvaptg}b) by first looking at ATMOMACCS alone. Among ATMOMACCS versions, performance improves from version 1 to 3: version 1 has similar MAEs to MACCS across all training set sizes, version 2 reduces MAE by 1.43 K, and version 3 further reduces it by 2.74 K. Version 4 continues this trend, lowering the error by an additional 0.58 K. In contrast, the other studied properties, version 5 performs worse than versions 3 and 4 by 1.07 K at the largest training set size (see Table \ref{tab:mae-comparison}). 

Compared to other descriptors, the topological fingerprint gives the highest MAE out of all descriptors (23.46 K). Meanwhile, MACCS and ATMOMACCS version 1 perform nearly identically across all training sizes (MAE 22.03 K). Finally, at the largest training size, ATMO reaches the same MAE as MACCS.

ATMOMACCS improves property predictions for most versions and properties, although the magnitude of improvement varied. Among all versions, ATMO-MACCS version 5 generally achieves the lowest prediction errors across the different property datasets. Accordingly, version 5 was selected for the feature importance analysis in Section \ref{sec:res_shap}.

\subsection{\label{sec:res_shap} SHAP analysis}
We now shift our focus to the inner workings of our descriptors to understand their relative performance in the different property prediction tasks. Our goal here is to qualitatively understand why ATMOMACCS works better than both ATMO and MACCS apart. Moreover, this section illustrates the interpretability of models trained using ATMOMACCS.

 We present the results of the SHAP analysis for ATMOMACCS version 5 separating datasets and target property prediction tasks, as above. Here, we have chosen to focus on ten features with the highest feature importance in terms of absolute value (taken both the negative and positive contributions). Next, the absolute feature importance values are grouped by category to provide a comprehensive overview. We chose to group the ATMOMACCS features as either belonging to non-oxygen counting MACCS keys, SIMPOL functional groups or motifs, carbon atom count, and oxygen atom count. 

\begin{figure}[!htbp]
\includegraphics[width=8.5cm]{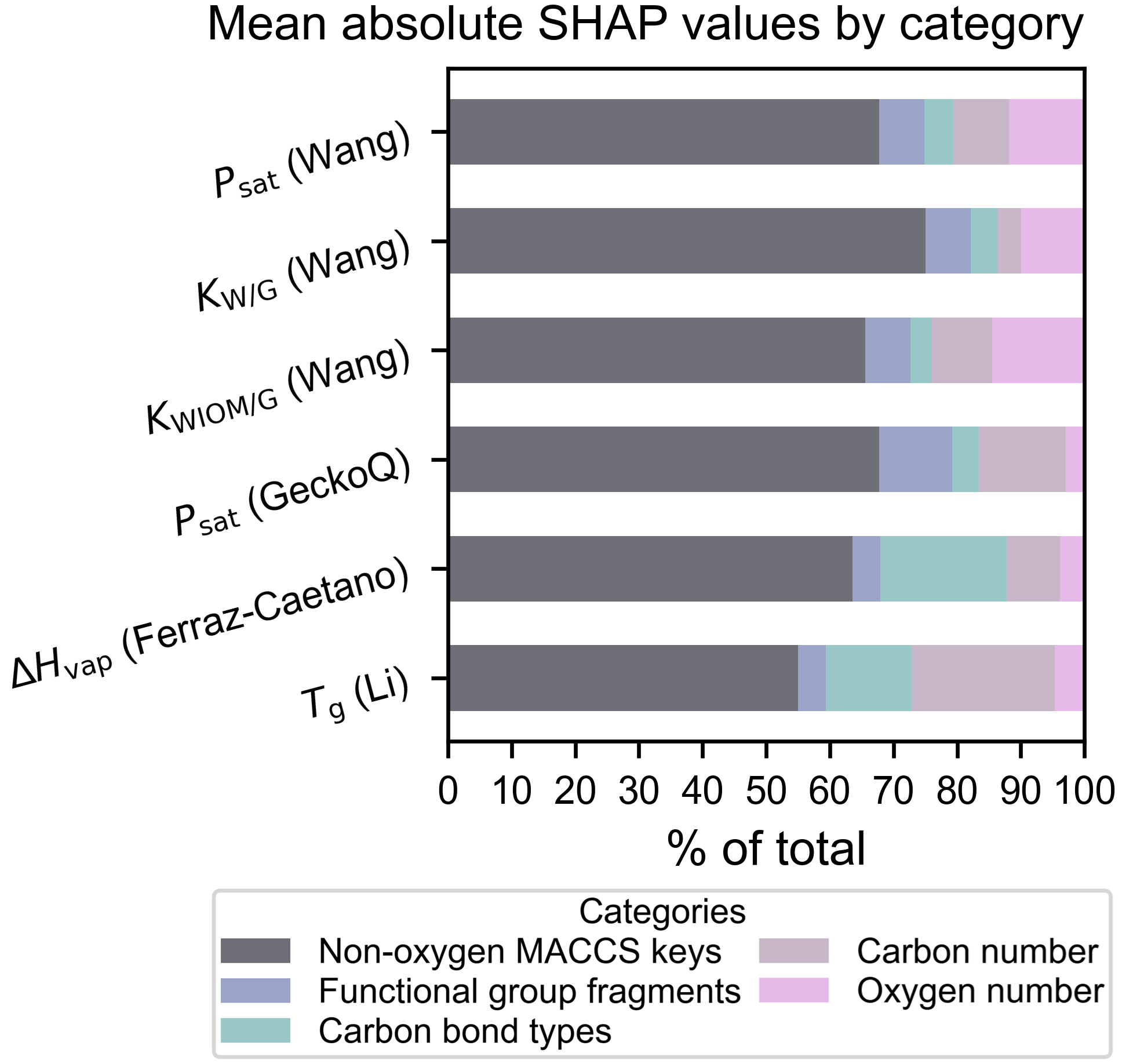}
\caption{ Mean absolute SHAP values aggregated by feature category for a kernel ridge regression (KRR) model based on ATMOMACCS version 5. The bars represent the importance of each feature category as a proportion of the total importance across all features. Acronyms:  P$_{sat}$ - saturation vapor pressure;  $K_{W/G}$ - water-gasphase equilibrium partition coefficient; $K_{WIOM/G}$ - water insoluble organic matter - gasphase equilibrium partition coefficient; $\Delta H_{vap}$ - enthalpy of vaporization; T$_g$ - glass transition temperature. Dataset names used in this work shown in parenthesis.}
\label{fig:shap_sum}
\end{figure}

Figure \ref{fig:shap_sum} shows the contribution of SHAP values by these categories. Despite MACCS performing worse in terms of prediction MAE shown in the previous sections, the combined importance of MACCS features is relatively large compared to ATMO features. The carbon bond types are more important for the \textit{Ferraz-Caetano} and \textit{Li} datasets.

\begin{figure*}[!htbp]
\includegraphics[width=17cm]{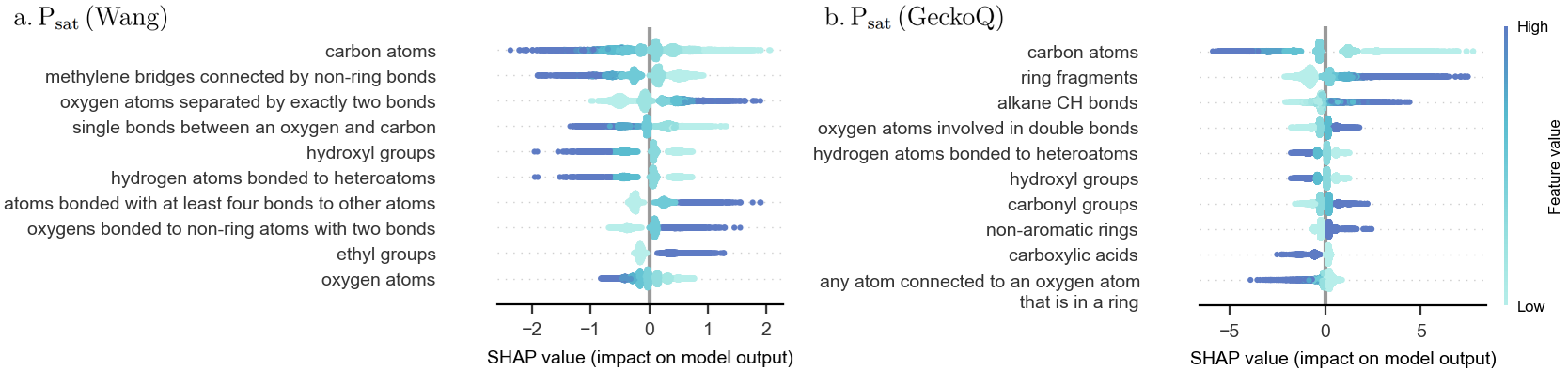}
\caption{Top features with largest absolute SHAP values for saturation vapor pressure $P_{sat}$ prediction in the (a) \textit{Wang} and (b) \textit{GeckoQ} dataset.}
\label{fig:shap_psat}
\end{figure*}

Figure \ref{fig:shap_psat}a shows the top ten most important features for $P_{sat}$ predictions on the \textit{Wang} dataset. The top features relate to carbon atom count, various (mostly carbon and oxygen related) bonding motifs, hydroxyl and ethyl groups, and oxygen atom counts. Figure \ref{fig:shap_psat}b shows the corresponding topmost important features for $P_{sat}$ prediction on the \textit{GeckoQ} dataset. Again, the carbon atom count is most important.  The oxygen atom count is ranked much less important for \textit{GeckoQ }than for \textit{Wang} in relative terms. 

\begin{figure*}[!htbp]
\includegraphics[width=17cm]{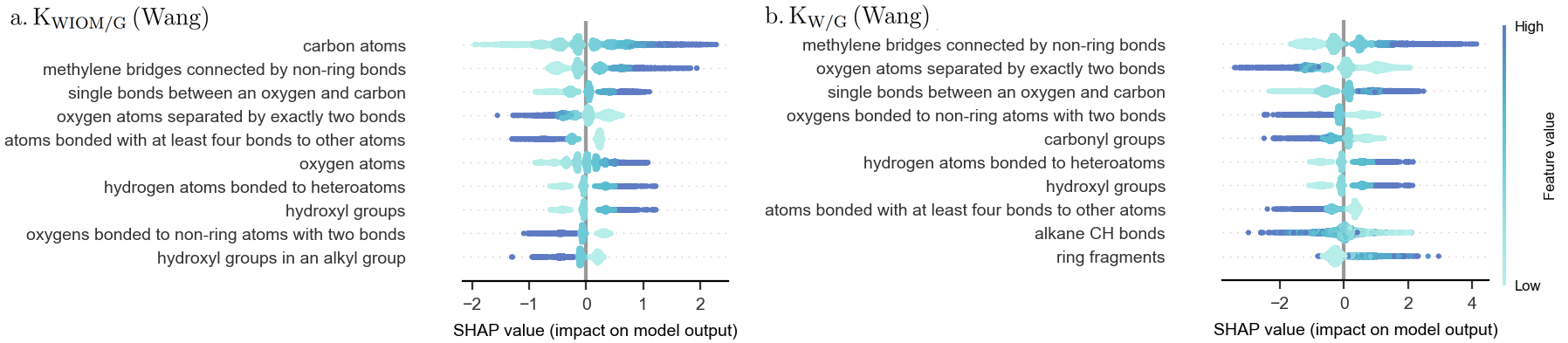}
\caption{The features with largest absolute SHAP values for equilibrium partition coefficient prediction in the  \textit{Wang}  dataset for partitioning between (a) the water insoluble organic matter-gasphase ($K_{WIOM/G}$) and (b) the water-gasphase ($K_{W/G}$).}
\label{fig:shap_kwg_wang}
\end{figure*}

The top features for both equilibrium partition coefficients are quite similar to $P_{sat}$ results on the \textit{Wang} dataset (Figure \ref{fig:shap_kwg_wang}). Especially for $K_{WIOM/G}$, the only difference compared to $P_{sat}$ is that hydroxyl groups in the context of an alkyl group were more important than the general hydroxyl group category. Notwithstanding this difference, $P_{sat}$ and $K_{WIOM/G}$ share top features, although their ranking differs slightly. Notably, the total number of carbon atoms and methylene bridges connected by non-ring bonds are the two most important features for predicting these properties. In contrast, for $K_{W/G}$ predictions, the number of carbon atoms is ranked 22nd, indicating much lower importance. For $ \log_{10}K_{W/G}$ predictions, the most important features involve bonding patterns around oxygen atoms, with carbon bonding topology playing a secondary role.

\begin{figure*}[!htbp]
\includegraphics[width=17cm]{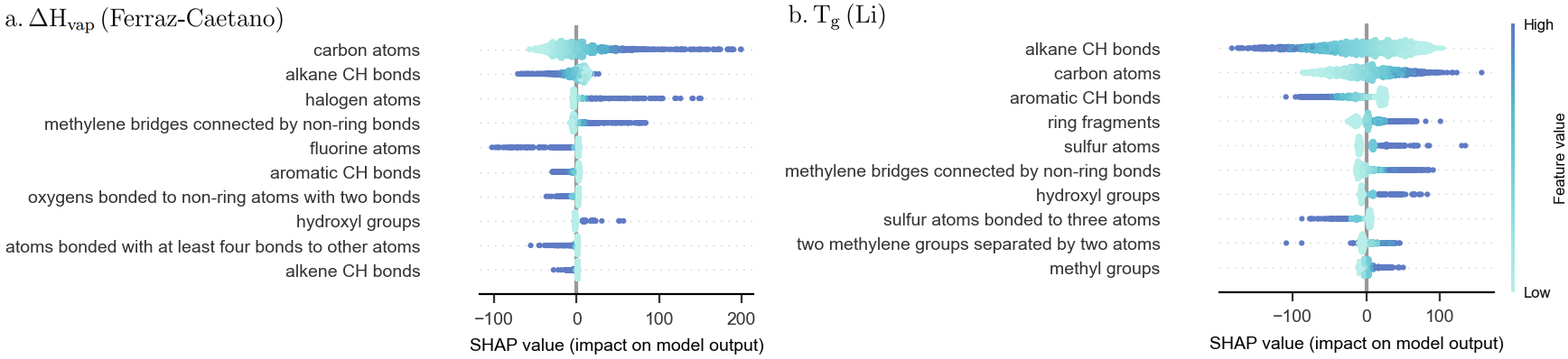}
\caption{The features with largest absolute SHAP values for (a) vaporization enthalpy ($\Delta H_{vap}$ ) and (b) glass transition temperature ($T_g$) prediction in the \textit{Ferraz-Caetano} and \textit{Li} datasets respectively.}
\label{fig:shap_fc_li}
\end{figure*}

Next, we examine the key characteristics to predict the $H_{vap}$ and the $T_{g}$ values of the molecules in the \textit{Ferraz-Caetano} and \textit{Li} datasets (Figure \ref{fig:shap_fc_li}a and \ref{fig:shap_fc_li}b, respectively). For these properties, many of the most important features relate to carbon hybridization and bonding topology, such as alkane CH, alkene CH, and aromatic CH. In contrast to the other properties, motifs related to oxygen bonding topology are not among the most influential.  Meanwhile, features associated with other elements, such as fluorine (\textit{Ferraz-Caetano}) and sulfur (\textit{Li}),  become important for these predictions. Nevertheless, the number of carbon atoms remains the top feature, ranking first and second for $\Delta H_{vap}$ and $T_{g}$, respectively. Methylene bridges and hydroxyl groups also appear among the top ten features.

\section{\label{sec:dis} Discussions\protect\\}
Our results in the previous section show that ATMOMACCS consistently outperforms the original MACCS fingerprint in the tested property prediction tasks (lower MAE), although the magnitude of improvement varies between versions and tasks. As shown in Table~\ref{tab:mae-comparison}, version~5, which maps ATMO and MACCS keys to integer counts, gives the lowest test set errors overall and is our recommended form of ATMOMACCS. Version~4, which encodes the same information as version~5 in binary form, performs slightly worse overall but is more memory-efficient for large-scale applications.

%ATMO PERFORMANCE
When used alone, ATMO performs similarly to MACCS, indicating that the combination of ATMO and MACCS is key to improved accuracy. MACCS features encode bonding topology and molecular connectivity, complementing the domain knowledge in the ATMO groups.

%ATMO AND ATMOMACCS VS SIMPOL
Figure \ref{fig:lc_psat_w_simpol} shows that ATMO and ATMOMACCS also improve upon the SIMPOL method from which the ATMO features were derived. A key factor contributing to ATMOMACCS’s improved performance, beyond the complementary MACCS features, is the nonlinear dependencies captured through the KRR model, compared to SIMPOL's linear form. ATMOMACCS allows the model to retain SIMPOL’s interpretability while achieving lower prediction errors. While our machine learning models were trained on compounds more closely related to the test sets than those used for SIMPOL parametrization, the pronounced error reduction nonetheless highlights the advantage of combining ATMO functional group knowledge with MACCS structural features and machine learning.

%PSAT ATMOMACCS TRENDS
We can further inspect which ATMOMACCS development steps were most effective by looking at version improvements (Figures \ref{fig:lc_psat}, \ref{fig:lc_kwq}, \ref{fig:lc_dvaptg}). For $P_{sat}$, adding higher SIMPOL motif counts (version~2) and the explicit carbon number (version~3) provides the largest improvements. Including oxygen counts (version~4) has only a limited effect, likely because oxygen-containing groups are already represented among other MACCS features. Figure \ref{fig:l_curves_appendix} shows that the influence of oxygen count does not increase when carbon number is excluded. This indicates that the oxygen count cannot replace the information provided by the carbon number. Together, these results suggest that carbon number in ATMOMACCS does not act primarily as an indicator of molecular size (as O:C ratio is close to 1 in both  \textit{Wang} and \textit{Gecko} molecules\cite{Sandstroem2025}) but encodes additional structural information relevant for property prediction.

% EQ CONST AND TG ATMOMACCS TRENDS
For the equilibrium constants $K_{W/G}$ and $K_{WIOM/G}$, performance improves most when higher SIMPOL motif counts are included (versions~2 and 5). Prediction of $K_{WIOM/G}$ also benefits from carbon count in version~3, whereas $K_{W/G}$ does not. The trends in $P_{sat}$ align more closely with those of $K_{WIOM/G}$, which is consistent with $P_{sat}$ representing the pure liquid--gas equilibrium, suggesting that atmospheric organics behave like water-insoluble compounds with extended nonpolar structures. ATMOMACCS performance trends for $T_g$ are similar to those for $P_{sat}$ and $K_{WIOM/G}$, indicating that these properties share common structural influences.

%HVAP ATMOMACCS TRENDS
For $H_{vap}$, carbon number has the strongest influence, reflecting the hydrocarbon-dominated nature of this dataset. This also explains why the standalone ATMO descriptor ranks highest for $H_{vap}$ and the \textit{Ferraz-Caetano} dataset, where differences among compounds are driven primarily by carbon number and backbone rather than functional group count or interactions (Figure \ref{fig:size_dist_appendix}). Interestingly, although $H_{vap}$ and $P_{sat}$ are fundamentally related, their descriptor ranking trends differ, likely reflecting differences in chemical composition between \textit{Ferraz-Caetano} and \textit{Wang}/\textit{GeckoQ}, which complicates direct comparisons.

%SHAP AND FEATURE GROUP CONTRIBUTIONS
Figure \ref{fig:shap_sum} highlights grouped SHAP contributions of ATMO and MACCS features. MACCS features consistently contribute more than 50\% of the total, showing their continued importance. Carbon and oxygen numbers are the next most influential features. Although comparisons of ATMOMACCS versions suggested only a small effect of oxygen count, SHAP analysis shows that oxygen-related motifs contribute approximately 5--15\% of total importance. This indicates that the small effect of oxygen count in ATMO is likely because MACCS already captures similar oxygen-related information. Retaining oxygen count in ATMOMACCS ensures that this complementary structural information is explicitly represented, which may benefit generalization across datasets.

%SHAP AND PROPERTY SPECIFICS
Feature analysis also shows that  $P_{sat}$ or partition coefficient predictions are primarily governed by carbon number and oxygen-related features. In contrast, $H_{vap}$ and $T_g$ are more sensitive to carbon--hydrogen bond types and heteroatoms other than oxygen. Although these two classes of properties are distinguished here, all ultimately contribute to the partitioning behavior of atmospheric compounds.

%SHAP AND SIMPOL
SHAP analysis further reveals the mechanistic basis of ATMOMACCS compared to SIMPOL. High SHAP magnitudes correspond to SIMPOL-derived motifs and carbon number, confirming their strong influence on predictions. Among the SIMPOL motifs, hydroxyl, carboxylic acid, hydroperoxide, and ketone are most influential for $P_{sat}$ prediction. These same features rank 3rd, 9th, 5th, and 15th, respectively, among SIMPOL’s fitted contributions.\cite{Pankow2008} Amides and amines are important in SIMPOL but are absent from our $P_{sat}$ datasets because such compounds are typically excluded from atmospheric mechanism simulations (e.g., MCM, GECKO-A) due to clustering behavior. Overall, combining MACCS and ATMO features with KRR enhances the predictive relevance of SIMPOL groups across datasets.

%COMPARISON TO LITERATURE
ATMOMACCS consistently outperforms the topological fingerprint and previously reported MAEs (Table \ref{tab:mae-comparison}). For example, we surpass the best-performing $P_{sat}$ model reported by Lumiaro \textit{et al.}\cite{Lumiaro2021}, which used a three-dimensional many-body tensor representation (MBTR) descriptor, reducing the MAE from 0.30 to 0.28~log$_{10}(P_{kPa})$ on the \textit{Wang} dataset. Similarly, Besel \textit{et al.}\cite{Besel2023,Besel2024} applied topological fingerprints with Gaussian process regression (GPR) to the \textit{GeckoQ} dataset, achieving an MAE of 0.82~log$_{10}(P_{kPa})$ on 3,637 test compounds, while Kr\"uger \textit{et al.}\cite{Kruger2025preprint} combined SIMPOL features with a graph neural network (GNN) to achieve 0.74~log$_{10}(P_{kPa})$. In comparison, our ATMOMACCS-based KRR model reaches 0.70~log$_{10}(P_{kPa})$, indicating that the descriptor efficiently captures the relevant molecular features despite the simpler model architecture.

%COMPARISON TO LIT CONTINUED.
For other properties, ATMOMACCS-KRR also outperforms previously reported models. Lumiaro \textit{et al.}\cite{Lumiaro2021} reported MAEs of 0.43 and 0.28 for $K_{W/G}$ and $K_{WIOM/G}$, compared to 0.39 and 0.26 with ATMOMACCS. \textit{Ferraz-Caetano et al.}\cite{FerrazCaetano2024} obtained 3.02~kJ~mol$^{-1}$ for $\Delta H_{\mathrm{vap}}$, whereas ATMOMACCS reduces this to 2.43~kJ~mol$^{-1}$. For \textit{Li et al.}\cite{Li2020} on $T_g$, direct comparison is limited due to missing MAE values.

%DATASET DIFF GECKOQ WANG
Consistent with earlier studies,~\cite{Besel2024} test set MAEs for \textit{GeckoQ} are roughly twice those of \textit{Wang} for all models, reflecting the greater molecular size and functional complexity of \textit{GeckoQ} compounds (Figures~\ref{fig:size_dist_appendix},~\ref{fig:fun_groups}). Dataset differences, including a 10~K temperature offset and a tenfold variation in size, prevent unbiased cross-dataset testing.

%SIMPOL PREDICTING OTHER PROPERTIES
Overall, these results suggest that ATMOMACCS effectively encodes molecular information relevant for multiple thermodynamic and physicochemical properties, outperforming both conventional fingerprints and more complex descriptors while remaining computationally efficient and interpretable. To our knowledge, this is the first demonstration that SIMPOL motifs contribute effectively to the prediction of $T_g$, $K_{WIOM/G}$ and $K_{W/G}$. This supports ATMOMACCS as a general-purpose descriptor for atmospheric organic compounds.

%LIMITATION OUTLOOK 
Despite these improvements, ATMOMACCS is currently limited to organic molecules, and caution is warranted when extrapolating beyond the training domain. Extending it to non-covalently bound systems such as clusters and aerosols is an important next step.\cite{Knattrup2023,Alfaouri2022,Tikkanen2022,Elm2020,Almeida2013} Beyond property prediction, ATMOMACCS could support unsupervised applications such as clustering atmospheric compounds or identifying compositional patterns in field data, thereby informing mechanistic modeling. Benchmarking against three-dimensional descriptors and GNNs can further test robustness. Still, interpretability and computational efficiency remain key advantages as larger datasets become available. In addition, ATMOMACCS is fully compatible with the open-source cheminformatics toolkit RDKit, enabling immediate use by the wider research community for molecular property prediction and descriptor generation.

%SUMMARY
In summary, chemically informed fingerprints such as ATMOMACCS enhance predictive accuracy, interpretability, and mechanistic understanding of atmospheric organic compounds. These results demonstrate the value of integrating interpretable chemical knowledge with machine learning for improved modeling of atmospheric processes.

\section{\label{sec:con} Conclusions\protect\\}
In conclusion, this study set out to investigate how molecular fingerprints can be optimized to better represent atmospheric organic compounds for property prediction using ATMOMACCS. By analyzing the performance of machine learning models across different ATMOMACCS versions, we found that incorporating functional groups and motifs specific to atmospheric chemistry markedly improves predictive accuracy compared to conventional fingerprints and group-contribution methods, while maintaining computational efficiency. We also found that integer-based feature encoding provides the best overall performance, although binary encodings perform only marginally worse, allowing users to select the appropriate version depending on dataset size and computational constraints. These findings advance our understanding of how molecular structure and representation influence the prediction of atmospheric compound properties. Moreover, ATMOMACCS shows strong potential as a molecular descriptor for large-scale atmospheric modeling. Future research could extend the ATMOMACCS framework to non-covalent systems and benchmark its performance against advanced molecular descriptors and neural network approaches.

\begin{acknowledgments}
This study was supported by the Academy of Finland through Project No. 346377 and the EU COST Actions CA18234 and CA22154. We further acknowledge CSC-IT Center for Science, Finland and the Aalto Science-IT project. The authors wish to acknowledge Theo Kurtén for insightful discussions. 
\end{acknowledgments}

\section*{Data Availability Statement}
The data and source code supporting the findings of this study are openly available on Zenodo at \url{https://doi.org/10.5281/zenodo.17231684}, reference number 17231684. The source code and data generated in this study are released under the Creative Commons Attribution-ShareAlike 4.0 International License (CC BY-SA 4.0). The datasets obtained from previous studies are republished in the Zenodo archive under their original licenses, preserving the terms and conditions specified by the original authors. Detailed licensing information for each dataset is provided in the accompanying Zenodo repository.

%\nocite{*}
\bibliography{01_bibliography}% Produces the bibliography via BibTeX.

\appendix

\section{Model Performance Summary, Data Stats, and Effect of Oxygen–Carbon Count Order}

\begin{table*}[htbp]
\centering
\renewcommand{\arraystretch}{1.1}
\setlength{\tabcolsep}{8pt}
\caption{The average mean absolute error (MAE) for our kernel ridge regression (KRR) model with all tested descriptors for all property prediction tasks at the largest training set size. Acronyms:  $P_{sat}$ - saturation vapor pressure;  $K_{W/G}$ - water-gasphase equilibrium partition coefficient; $K_{WIOM/G}$ - water insoluble organic matter - gasphase equilibrium partition coefficient; $\Delta H_{vap}$ - enthalpy of vaporization; $T_g$ - glass transition temperature. Error reduction represents the percentual decrease in mean absolute error (MAE) of ATMOMACCS v5 compared with the topological fingerprint. Values are rounded to the nearest integer for clarity.}
\begin{tabular}{l r r r r r r}

Descriptor & \textit{Wang} $P_{sat}$ & \textit{Wang} $K_{W/G}$ & \textit{Wang} $K_{WIOM/G}$ & \textit{GeckoQ} $P_{sat}$ & \textit{Li} $T_{g}$ & \textit{Ferraz-Caetano} $\Delta H_{vap}$ \\
\hline
MACCS fingerprint   & 0.44 & 0.52 & 0.41 & 1.11 & 22.03 & 10.10 \\
ATMO v4             & 0.45 & 0.65 & 0.41 & 0.84 & 24.15 & 5.36 \\
ATMO v5             & 0.43 & 0.61 & 0.38 & 0.82 & 22.18 & 5.03 \\
ATMOMACCS v1        & 0.39 & 0.46 & 0.37 & 0.97 & 21.98 & 10.10 \\
ATMOMACCS v2        & 0.34 & 0.43 & 0.31 & 0.78 & 20.56 & 10.02 \\
ATMOMACCS v3        & 0.30 & 0.43 & 0.27 & 0.73 & 17.82 & 2.81 \\
ATMOMACCS v4        & 0.29 & 0.42 & 0.27 & 0.73 & 17.24 & 2.82 \\
ATMOMACCS v5        & 0.28 & 0.39 & 0.26 & 0.70 & 18.31 & 2.43 \\
Topological fingerprint & 0.31 & 0.41 & 0.29 & 0.75 & 23.46 & 6.29 \\
Error reduction ($\%$) & 8 & 5 & 9 & 7 & 22 & 61 \\
\hline
\end{tabular}
\label{tab:mae-comparison}
\end{table*}

\begin{figure*}
    \centering
    \includegraphics[width=17cm]{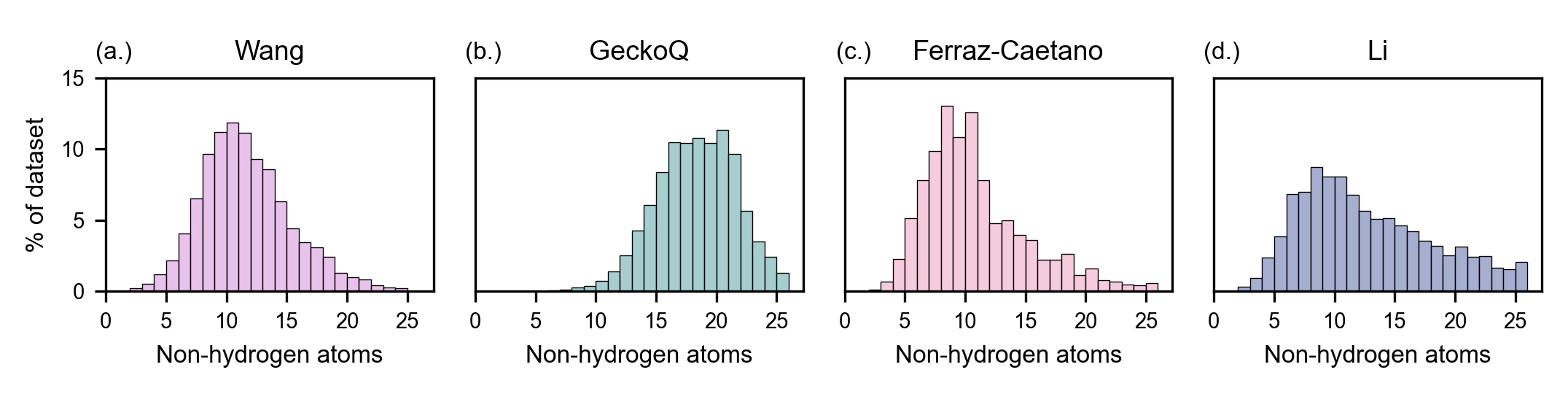}
    \caption{Size distributions for the four datasets considered. The molecule size in terms of mass is largely determined by the number of non-hydrogen atoms present.}
    \label{fig:size_dist_appendix}
\end{figure*}

\begin{figure*}[]
\includegraphics[width=17cm]{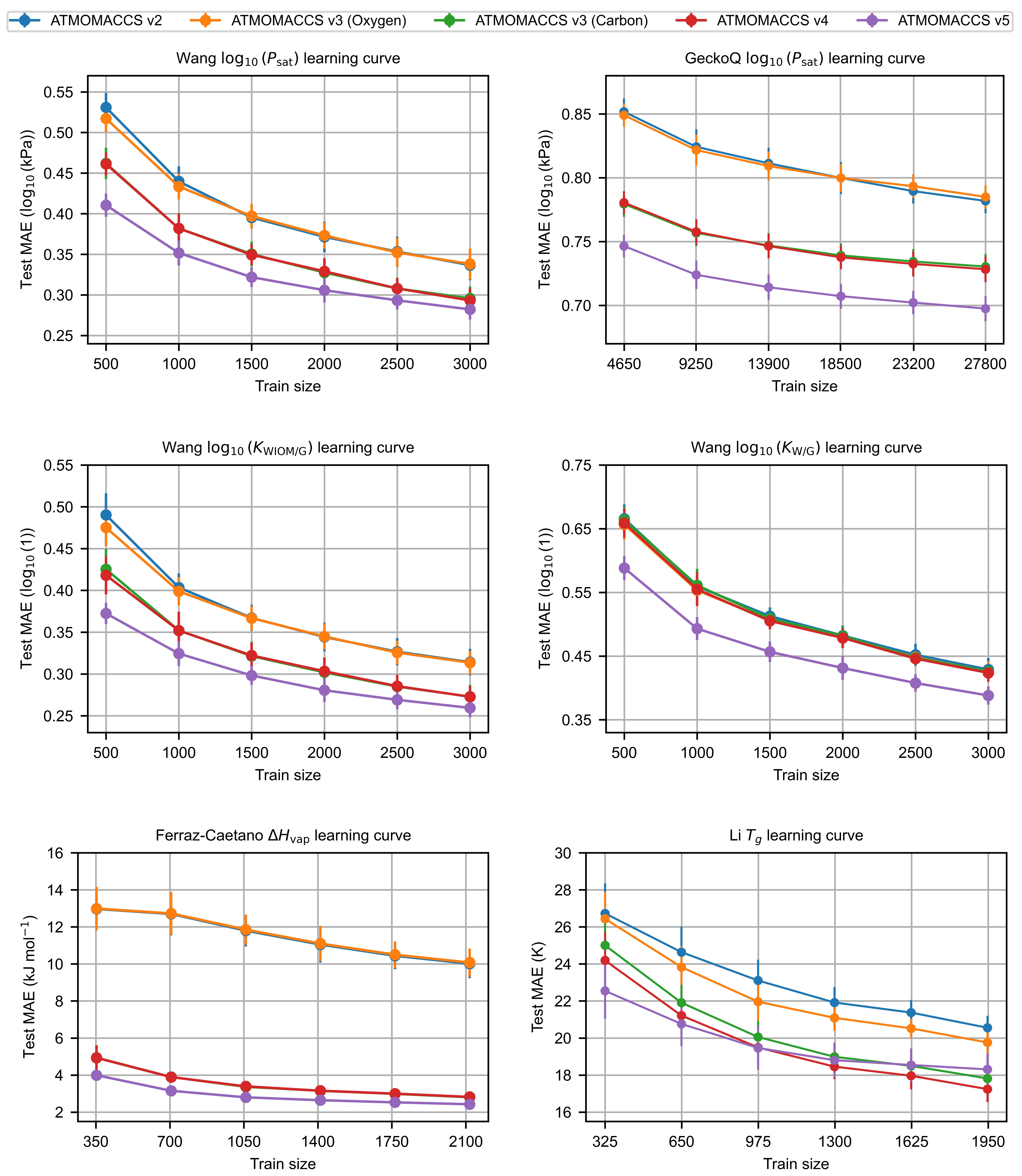}
\caption{Learning curves including alternative ATMOMACCS version 3 (Oxygen) version. Acronyms:  P$_{sat}$ - saturation vapor pressure;  $K_{W/G}$ - water-gasphase equilibrium partition coefficient; $K_{WIOM/G}$ - water insoluble organic matter - gasphase equilibrium partition coefficient; $\Delta H_{vap}$ - enthalpy of vaporization; T$_g$ - glass transition temperature. Dataset names assoiciated with each target are found in panel titles.}
\label{fig:l_curves_appendix}
\end{figure*}

\end{document}